\documentclass[aps,pra,superscriptaddress,showpacs,showkeys]{revtex4}
\usepackage{pstricks}
\usepackage{epsfig}
\usepackage{graphicx}
\usepackage{amsmath}
\usepackage{amsfonts}
\usepackage{amssymb}
\usepackage{wasysym}
\usepackage{bbold}
\usepackage{rotating}
\newcommand{\bra}[1]{\langle #1|}
\newcommand{\ket}[1]{|#1\rangle}
\newcommand{\braket}[2]{\langle #1|#2\rangle}
\begin{document}
\title{Variational determination of the second-order density matrix for the isoelectronic series of Beryllium, Neon and Silicon}
\author{Brecht Verstichel}
\email{brecht.verstichel@ugent.be}
\affiliation{Center for Molecular Modeling, Ghent University, Proeftuinstraat 86, 9000 Gent, Belgium}
\author{Helen van Aggelen}
\affiliation{Department of Inorganic and Physical Chemistry, Ghent University, Krijgslaan 281 (S3), 9000 Gent, Belgium}
\author{Dimitri Van Neck}
\affiliation{Center for Molecular Modeling, Ghent University, Proeftuinstraat 86, 9000 Gent, Belgium}
\author{Paul W. Ayers}
\affiliation{Department of Chemistry, McMaster University, Hamilton, Ontario, L8S 4M1, Canada }
\author{Patrick Bultinck}
\affiliation{Department of Inorganic and Physical Chemistry, Ghent University, Krijgslaan 281 (S3), 9000 Gent, Belgium}
\begin{abstract}
The isoelectronic series of Be, Ne and Si are investigated using a variational determination of the second-order density matrix. A semidefinite program was developed that exploits all rotational and spin symmetries in the atomic system. We find that the method is capable of describing the strong static electron correlations due to the incipient degeneracy in the hydrogenic spectrum for increasing central charge. Apart from the ground-state energy various other properties are extracted from the variationally determined second-order density matrix. The ionization energy is constructed using the extended Koopmans' theorem. The natural occupations are also studied, as well as the correlated hartree-Fock-like single particle energies. The exploitation of symmetry allows to study the basis set dependence and results are presented for correlation-consistent polarized valence double, triple and quadruple zeta basis sets.
\end{abstract}
\pacs{31.15.-p,31.15.A-,31.15.xt}
\keywords{isoelectronic series, variational method, density matrix, semidefinite program}
\maketitle
\section{Introduction}
The idea of a variational determination of  the ground-state energy for a nonrelativistic many-body problem based
on the second-order density matrix (2DM) has a long history \cite{lowdin,coleman,garrod} and several highly appealing features.
The energy of a system is a known linear functional of the 2DM. $N$-particle wave functions never
need to be manipulated since the energy is minimized directly in terms of the 2DM.
However, the minimization is constrained because the  variational search should be done
exclusively with 2DMs that can be derived from an $N$-particle wave function (or an ensemble of $N$-particle wave functions).
Such a 2DM is called $N$-representable, and the complexity of the many-body problem is in fact shifted to the
characterization of this set of $N$-representable 2DMs.
The complete (necessary and sufficient) set of conditions for $N$-representability of a 2DM is not known in a constructive form, but it is clear that the energy from a minimization constrained by a set of necessary $N$-representability conditions is a strict lower bound to the exact energy. Therefore this approach is highly complementary to the usual variational procedure based on a wave-function ansatz, which produces upper bounds.
In addition the method is in principle exact, in the sense that as increasingly accurate set of $N$-representability conditions are imposed in the minimization, the resulting energy converges to the exact one.

These are fascinating ideas for any true-blooded many-body theorist, as it comes close to the ``ultimate reduction''
of an interacting many-particle problem to solving a sequence of two-particle problems. In practice, however,
implementing the method turns out to be very difficult and it is only in the last decade that serious attempts have been undertaken to turn the idea into a practical calculational scheme. The massive efforts by Mazziotti et al. \cite{mazziotti,hammond,maz_prl} and Nakata et al. \cite{nakata_first,nakata_last} are particularly notable. The main difficulty is of a technical nature: stringent $N$-representability conditions require the positive semidefiniteness of matrix functionals of the 2DM, which turns the variational problem into a so-called semidefinite program (SDP).
Even applying the simplest ``two-index'' conditions, a direct energy minimization using Newton-Raphson methods requires
a matrix operation scaling as $M^{12}$ (where $M$ is the number of single-particle states) in each Newton-Raphson step.
This can be circumvented in various ways, so that only matrix operations scaling as $M^6$ are needed.
While these are \textit{nominally} $M^6$ methods, the number of iterations required to reach convergence is
very high and seems to rise with system size; in practice present implementations are probably about 100-1000 times slower than comparable methods such as CCSD. Still, one has the feeling that there is potential to turn it into a \textit{genuine} $M^6$ method, and it is of interest to investigate the properties of SDP applied to various  systems.

Up to now, most applications covered electronic structure calculations in atoms and molecules. Attention has been given
primarily to the resulting energy. In this paper, we focus on three issues: (i) the performance of SDP in
multireference situations (strong static correlations), (ii) the quality of the variationally obtained 2DM, and (iii) the dependence
of the results on $M$ (the size of the basis set).
We do this by investigating three well-known examples in electronic structure theory: the isoelectronic series of Be, Ne, and Si.
It is well known that the correlation energy for $N$ electrons in the field of a positive point charge $Z$ has a $Z$-dependence that strongly depends on $N$. For an increasing central charge $Z$, the hartree-Fock spectrum tends to
the hydrogenic one, which has an ``accidental'' degeneracy related to a special symmetry in the Coulomb Hamiltonian.
For the four-electron series the incipient degeneracy of the $2p$ and $2s$ orbitals leads to a vanishing particle-hole gap, inducing
strong correlation effects with a correlation energy proportional to $Z$. For the ten-electron series this does not happen because a major shell is closed, and the correlation energy becomes flat for increasing $Z$. The 14-electron series again shows degeneracy effects, and in addition is a spin triplet.

In Sec.~II we provide theoretical and calculational backgrounds on the SDP implementation that is used. (Note that the techniques developed in this section have been used previously to study molecular dissociation \cite{helen_1,helen_2}.) In particular, we pay attention to the way spin and rotational symmetry are exploited, enabling the use of quite large (cc-pVQZ) basis sets. In Sec.~III the SDP results for the isolectronic series of Be, Ne, and Si are discussed. A summary is provided in Sec.~IV. Atomic units are used throughout the paper.
\section{Theory}

\subsection{$N$-representability conditions}

We will use second quantized notation where $a_\alpha^\dagger$ ($a_\alpha$) creates (annihilates) an electron in a single-particle 
(sp) state $\alpha$. The Hamiltonian can be written as
\begin{eqnarray}
\hat{H} &=& \sum_{\alpha\gamma}t_{\alpha\gamma}a^\dagger_\alpha a_\gamma + \frac{1}{4}\sum_{\alpha\beta\gamma\delta}V_{\alpha\beta;\gamma\delta}a^\dagger_\alpha a^\dagger_\beta a_\delta a_\gamma~,
\end{eqnarray}
where $t_{\alpha\gamma}$ is the matrix element of the one-body part of the Hamiltonian (kinetic energy plus external potential)
 and $V_{\alpha\beta;\gamma\delta}$ is the antisymmetrized matrix element of the Coulomb interaction.
The problem of finding the ground state of a quantum mechanical many-body system can be reformulated in terms of the second-order 
density matrix
\begin{equation}
\Gamma_{\alpha\beta;\gamma\delta} = \bra{\Psi^N}a^\dagger_\alpha a^\dagger_\beta a_\delta a_\gamma\ket{\Psi^N}~.
\end{equation}
In principle $\Gamma$ is a complex Hermitian matrix, but for a Coulomb Hamiltonian it is sufficient to consider 
real-symmetric matrices,
\begin{equation}
\Gamma_{\alpha\beta;\gamma\delta} = \Gamma_{\gamma\delta;\alpha\beta}~.
\end{equation}
In addition $\Gamma$ obeys the fermionic relations for antisymmetry in the sp indices
\begin{equation}
\Gamma_{\alpha\beta;\gamma\delta} = -\Gamma_{\beta\alpha;\gamma\delta} = -\Gamma_{\alpha\beta;\delta\gamma} = \Gamma_{\beta\alpha;\delta\gamma}~.
\end{equation}
The density matrix $\Gamma$ can be determined variationally through the minimization of the energy functional
\begin{equation}
E(\Gamma) = \mathrm{Tr}~\left[\Gamma H^{(2)}\right] = \frac{1}{4} \sum_{\alpha\beta;\gamma\delta}\Gamma_{\alpha\beta;\gamma\delta}H^{(2)}_{\alpha\beta;\gamma\delta}~,
\label{tp_ham}
\end{equation}
where the reduced two-particle (tp) Hamiltonian is defined as
\begin{eqnarray}
\nonumber H^{(2)}_{\alpha\beta;\gamma\delta} &=& \frac{1}{N - 1}\left(t_{\alpha\gamma}\delta_{\beta\delta} - 
t_{\alpha\delta}\delta_{\beta\gamma} - t_{\beta\gamma}\delta_{\alpha\delta} + t_{\beta\delta}\delta_{\alpha\gamma}\right) + V_{\alpha\beta;\gamma\delta}~.
\end{eqnarray}
The problem with this method is that the complete set of conditions that the density matrix has to fulfill to be derivable 
from a physical wave function (the so-called $N$-representability conditions) is not known in a constructive form \cite{payers}. 
Therefore one minimizes the energy functional under a limited set of $N$-representability conditions. Three simple conditions, known as the $P$, $Q$ and $G$ conditions \cite{coleman,garrod}, are known to give quite good results. The $P$ condition expresses the fact 
that the 2DM has to be positive semidefinite.
The physical interpretation of the $Q$ condition is that the two-hole matrix, $Q$, has to be positive semidefinite; 
using basis anticommutation relations $Q$ can be written as a homogeneous linear mapping, from the tp matrix space onto itself:
\begin{eqnarray}
\nonumber Q_{\alpha\beta;\gamma\delta} &=& \bra{\Psi^N}a_\alpha a_\beta a^\dagger_\delta a^\dagger_\gamma\ket{\Psi^N}\\
&=&\Gamma_{\alpha\beta;\gamma\delta} + \frac{1}{n}\left(\delta_{\alpha\gamma} \delta_{\beta\delta} - \delta_{\alpha\delta} \delta_{\beta\gamma}\right)\mathrm{Tr}~\Gamma -\delta_{\beta\delta}\rho_{\alpha\gamma} + \delta_{\alpha\delta }\rho_{\beta\gamma} -  
\delta_{\alpha\gamma}\rho_{\beta\delta} +  \delta_{\beta\gamma}\rho_{\alpha\delta}~.
\end{eqnarray}
Here the particle number constraint has been used
\[\mathrm{Tr}~\Gamma = \frac{N(N-1)}{2} = n~,\]
as well as the definition of the sp density matrix:
\begin{equation}
\rho_{\alpha\gamma} = \frac{1}{N-1}\sum_\beta \Gamma_{\alpha\beta;\gamma\beta}~.
\end{equation}
The $G$ condition demands that the particle-hole (ph) matrix $G$, is positive semidefinite; again, $G$ can be written as a
 homogeneous linear mapping, from the tp matrix space onto the ph matrix space:
\begin{eqnarray}
\nonumber G_{\alpha\beta;\gamma\delta} &=& \bra{\Psi^N}a^\dagger_\alpha a_\beta a^\dagger_\delta a_\gamma \ket{\Psi^N}\\
&=& \delta_{\beta\delta}\rho_{\alpha\gamma} - \Gamma_{\alpha\delta;\gamma\beta}~.
\end{eqnarray}
Recently there has been progress on improved $N$-representability conditions using the positive semidefiniteness of higher-order density matrices, e.g. the three-positivity conditions known as the $T_1$ and $T_2$ conditions \cite{zhao,hammond}. Also some attempts have been made to improve $N$-representability while remaining strictly in tp space, by considering Hamiltonian dependent positivity conditions \cite{maz_ham}, or sharp bounds on the $P$, $Q$ and $G$ operators \cite{dimi}. 
However, in the present paper we restrict ourselves to the standard $P$, $Q$ and $G$ conditions.

\subsection{Inclusion of spin symmetry}

\subsubsection{General case}

When the Hamiltonian of the system is invariant under rotations in spin space, the eigenstates can be characterized by their total spin $\Sigma$ and spin projection $\mu$. 
Explicitly introducing the electron spin, a sp state is written as $\{\ket{\alpha} \equiv \ket{as_a}\}$, where $a$ is the spatial 
orbital index and $s_a = \pm\frac{1}{2}$ is the spin projection. Two sp states can couple to a pair with total spin $S = 0$ or $S = 1$. 
The corresponding pair creation operator is
\begin{eqnarray}
B^\dagger_{ab;SM} &=& \left[a^\dagger_a \otimes a^\dagger_b\right]^S_M\\
&=& \sum_{s_as_b}\braket{ \tfrac{1}{2} s_a \tfrac{1}{2} s_b}{S M}a^\dagger_{as_a}a^\dagger_{bs_b}
\end{eqnarray}
and the density matrix $\Gamma$ in spin-coupled tp space is defined as
\begin{equation}
\label{spincoupled_rdm}
^{{\Sigma\mu}}\Gamma^{SM;S'M'}_{ab;cd} = \bra{\Psi^N_{{\Sigma\mu}}}B^\dagger_{ab;SM} B_{cd;S'M'}\ket{\Psi^N_{{\Sigma\mu}}}~.
\end{equation}
The $B^\dagger B$ operator in Eq. (\ref{spincoupled_rdm}) can now be further coupled to an object with good total spin. 
First one has to introduce $\tilde{B}$,
\begin{equation}
\tilde{B}_{cd;SM} = (-1)^{S+M}B_{cd;S~-M},
\end{equation}
which is again a good spherical tensor operator. 
Equation (\ref{spincoupled_rdm}) can now be rewritten as,
\begin{eqnarray}
^{\Sigma\mu}\Gamma^{SM;S'M'}_{ab;cd} &=& (-1)^{S'-M'} \sum_{S_T}\braket{S~MS'-M'}{S_T 0} \bra{\Psi^N_{{\Sigma\mu}}}\left[B^\dagger_{ab;S}\otimes \tilde{B}_{cd;S'}\right]^{S_T}_{0}\ket{\Psi^N_{{\Sigma\mu}}}~.
\label{gamma_coupled}
\end{eqnarray}
The density matrices on the right of Eq.~(\ref{gamma_coupled}) are classified by $S_T = 0,1,2$ and provide an equivalent 
representation of the 2DM of the $\mu$th member of the spin multiplet. 
Note that the 2DMs of different members are trivially related through the Wigner-Eckart theorem, 
\begin{eqnarray}
\bra{\Psi^N_{{\Sigma\mu}}}\left[B^\dagger_{ab;S}\otimes \tilde{B}_{cd;S'}\right]^{S_T}_{0}\ket{\Psi^N_{{\Sigma\mu}}}&=& \frac{(-1)^{\Sigma-\mu}}{[S_T]}\braket{{\Sigma~\mu \Sigma-\mu}}{S_T 0}\bra{\Psi^N_{\Sigma}}|\left[B^\dagger_{ab;S}\otimes \tilde{B}_{cd;S'}\right]^{S_T}|\ket{\Psi^N_{\Sigma}}~,
\label{wigner-eckart}
\end{eqnarray}
in terms of reduced matrix elements. Here $[S] = \sqrt{2S+1}$.  

\subsubsection{Singlet ground state}

If the ground state has $\Sigma = 0$ (spin singlet), the number of
matrices involved in the minimization procedure is significantly reduced.
Obviously for a singlet ground state the operator in Eq. (\ref{gamma_coupled}) has to be scalar, i.e.,\ only the 
$S_T = 0$ part is nonzero, and Eq.~(\ref{gamma_coupled}) reduces to
\begin{eqnarray}
^{00}\Gamma^{SM;S'M'}_{ab;cd} &=&\delta_{SS'}\delta_{M M'}~ \Gamma^S_{ab;cd}
\end{eqnarray}
where 
\begin{eqnarray}
\Gamma^S_{ab;cd} = \bra{\Psi^N_{00}}B^\dagger_{ab;SM}B_{cd;SM}\ket{\Psi^N_{00}}~.
\end{eqnarray}
is independent of $M$. 
This shows that, for a singlet ground state, the density matrix in a coupled tp basis is diagonal in $S$ and $M$, and independent of the
spin projection $M$. Instead of having to work with the full density matrix, all matrix manipulations can be performed on only two diagonal 
blocks, the $S=0$ and $S=1$ matrices, which are respectively symmetric and antisymmetric in the indices related to the spatial orbitals. 

We now reformulate the minization problem in the spin-coupled representation. The $Q$-matrix in the coupled representation is similarly 
defined as:
\begin{equation}
Q^S_{ab;cd} = \bra{\Psi^N_{00}}{B}_{ab;SM} B^\dagger_{cd;SM}\ket{\Psi^N_{00}}~.
\end{equation}
It is clear that the $Q$ matrix has an identical block diagonal structure as $\Gamma$. After some recoupling one can 
write the $Q$-mapping, from coupled tp space onto coupled tp space, as 
\[Q^S_{ab;cd} = \Gamma^S_{ab;cd} + \frac{1}{n}(\delta_{ac}\delta_{bd} + (-1)^S\delta_{ad}\delta_{bc})\mathrm{Tr}~\Gamma - \rho_{ac}\delta_{bd} - (-1)^S \rho_{bc}\delta_{ad} - \rho_{bd}\delta_{ac} - (-1)^S\rho_{ad}\delta_{bc}~,\]
where the sp matrix $\rho$
\begin{eqnarray}
\rho_{ac} &=& \delta_{s_as_c}\bra{\Psi^N_{00}}a^\dagger_{as_a}a_{cs_a}\ket{\Psi^N_{00}}\\
&=& \frac{1}{2}\frac{1}{N-1}\sum_{S}\left[S\right]^2\sum_l\Gamma^S_{al;cl}~
\end{eqnarray}
and the trace
\begin{equation}
\mathrm{Tr}~\Gamma = \frac{1}{2}\sum_S [S]^2 \sum_{ab} \Gamma^S_{ab;ab}~,
\end{equation}
can be expressed in terms of the coupled $\Gamma^S$.

The $G$-matrix is a bit more involved. The coupled ph creation operator reads
\begin{eqnarray}
A^\dagger_{ab;SM} &=& \left[a^\dagger_a \otimes \tilde{a}_b\right]^S_M\\
&=&\sum_{s_as_b}(-1)^{\frac{1}{2} - s_b} \braket{\tfrac{1}{2} s_a \tfrac{1}{2} -s_b}{SM} a^\dagger_{as_a} a_{bsb}\nonumber~.
\end{eqnarray}
The $G$-matrix in coupled ph space can now be written as:
\begin{equation}
G^S_{ab;cd} = \bra{\Psi^N_{00}}A^\dagger_{ab;SM}A_{cd;SM}\ket{\Psi^N_{00}}~.
\end{equation}
Again, one can prove that this matrix has the same block structure as the $\Gamma$ and $Q$ matrices.
After some angular momentum recoupling we get the expression for the $G$-map in the coupled representation:
\begin{equation}
G^S_{ab;cd} = \delta_{bd}\rho_{ac} - \sum_{S'}\left[S'\right]^2\left\{\begin{matrix}\frac{1}{2} & \frac{1}{2} 
& S\\\frac{1}{2} & \frac{1}{2} & S'\end{matrix}\right\}\Gamma^{S'}_{ad;cb}~.
\end{equation}

\subsubsection{Nonsinglet states\label{spinensemble}}

For higher-spin multiplets the same block decomposition is possible, provided a spin-averaged ensemble is considered. 
The density matrix for such an ensemble is defined in spin-coupled representation as
\begin{eqnarray}
^\Sigma\Gamma^{SS';M}_{ab;cd} &=& \frac{1}{2\Sigma+1}\sum_{\mu}\bra{\Psi^N_{\Sigma\mu}}B^\dagger_{ab;SM}B_{cd;S'M}\ket{\Psi^N_{\Sigma\mu}}~.
\label{nonsinglet_rdm}
\end{eqnarray}
Note that the minimal energy can be reached for such an ensemble, since all members of the multiplet are degenerate. 
Performing the same manipulation as leading to Eq.~(\ref{gamma_coupled}) and using the Wigner-Eckart theorem as in Eq.~(\ref{wigner-eckart}) one obtains  
\begin{equation}
^\Sigma\Gamma^{SS';M}_{ab;cd} = \sum_{\mu}\frac{(-1)^{S'-M}}{[\Sigma]^2}\sum_{S_T} \braket{S~MS'-M}{S_T 0} \frac{(-1)^{{\Sigma-\mu}}}{[S_T]}\braket{{\Sigma~\mu \Sigma-\mu}}{S_T 0} \bra{\Psi^N_{\Sigma}}|\left[B^\dagger_{ab;S}\otimes \tilde{B}_{cd;S'}\right]^{S_T}|\ket{\Psi^N_{\Sigma}}~.
\label{nonsinglet1}
\end{equation}
Since $(-1)^{{\Sigma-\mu}}/[\Sigma] = \braket{{\Sigma \mu \Sigma-\mu}}{00}$ one can use 
orthogonality of the Clebsch-Gordan coefficients to work out 
the sum over $\mu$ in Eq.~(\ref{nonsinglet1}). The result 
\begin{eqnarray}
^\Sigma\Gamma^{SS';M}_{ab;cd} &=& \frac{\delta_{SS'}}{[S][\Sigma]}
\bra{\Psi^N_{\Sigma}}|\left[B^\dagger_{ab;S}\otimes \tilde{B}_{cd;S}\right]^{0}|\ket{\Psi^N_{\Sigma}}~,
\end{eqnarray} 
implies that the ensemble 2DM is again block-diagonal in spin, and the same formulas can be used as for the singlet case. 

\subsection{Inclusion of rotational symmetry}

In atomic systems, the rotational symmetry of the Hamiltonian further reduces the dimension of the blocks involved in the density matrix. In exactly the same way as for spin, one can show that the density matrix of an ensemble, when avaraged over the third component of angular momentum, is diagonal in the two-particle angular momentum $L$ and its $z$-component $M_L$, and completely independent of the value of $M_L$. What is more, for atomic systems there is also the parity ($\pi=\pm1$) of the two particle states.  In the end one gets a density matrix that is composed out of blocks with fixed values for $L^\pi S$, enabling one to solve the variational problem in large basis sets. The sp basis for systems with rotational and spin symmetry is written as $\ket{am_as_a}$, where $a$ is shorthand for the radial basis state $n_al_a$. The tp density matrix in spin and angular momentum coupled representation is defined as
\begin{equation}
\Gamma^{(L^\pi S)}_{ab;cd} = \bra{\Psi^N}B^\dagger_{ab;L^\pi S}B_{cd;L^\pi S}\ket{\Psi^N}~,
\end{equation}
where
\begin{eqnarray}
\nonumber B^\dagger_{ab;L^\pi S} &=& \left[a^\dagger_a \otimes a^\dagger_b\right]^{LS}_{M_LM_S}\\
&=&\sum_{m_am_b}\sum_{s_as_b} \braket{l_a m_a l_b m_b}{LM_L} \braket{\frac{1}{2} s_a \frac{1}{2} s_b}{SM_S} a^\dagger_{n_al_am_as_a}a^\dagger_{n_bl_bm_bs_b}~.
\end{eqnarray}
In an analogous way as for spin coupling, the spin and angular momentum coupled $Q$-matrix is defined as
\begin{equation}
Q^{(L^\pi S)}_{ab;cd} = \bra{\Psi^N}B_{ab;L^\pi S} B^\dagger_{cd;L^\pi S}\ket{\Psi^N}~,
\end{equation}
out of which the coupled $Q$-map can be derived
\begin{eqnarray}
\nonumber Q^{(L^\pi S)}_{ab;cd} &=& \Gamma^{(L^\pi S)}_{ab;cd} + \frac{\mathrm{Tr}~\Gamma}{n}\left(\delta_{ac}\delta_{bd} + (-1)^{L + S + l_c + l_d}\delta_{ad}\delta_{bc}\right)\\
&& -\delta_{bd}\delta_{l_al_c}\rho^{(l_a)}_{n_an_c} - \delta_{ac}\delta_{l_bl_d}\rho^{(l_b)}_{n_bn_d}~,
\end{eqnarray}
with the sp density matrix defined as
\begin{equation}
\rho^{(l_a)}_{n_an_c} = \frac{1}{2}\frac{1}{2l_a+1}\frac{1}{N-1}\sum_{(L^\pi S)}\left[L\right]^2\left[S\right]^2 \sum_{n_bl_b}\Gamma^{(L^\pi S)}_{n_al_an_bl_b;n_cl_an_bl_b}~.
\end{equation}
The $G$-matrix is defined as
\begin{equation}
G^{(L^\pi S)}_{ab;cd} = \bra{\Psi^N}\left[a^\dagger_a \otimes \tilde{a}_b\right]^{(L^\pi S)}\left(\left[a^\dagger_c\otimes \tilde{a}_d\right]^{(L^\pi S)}\right)^\dagger\ket{\Psi^N}~,
\end{equation}
where again $\tilde{a}$ is a spherical tensor operator defined as
\begin{equation}
\tilde{a}_{bm_bs_b} = (-1)^{l_b+m_b+\frac{1}{2} + s_b}a_{b-m_b-s_b}~.
\end{equation}
The spin and angular momentum coupled $G$-map from tp space on ph space becomes
\begin{equation}
G^{(L^\pi S)}_{ab;cd} = \delta_{bd}\delta_{l_al_c}\rho^{(l_a)}_{n_an_c} - \sum_{(L^\pi S)'}\left[S'\right]^2\left[L'\right]^2\left\{\begin{matrix}l_d & l_c & L\\l_b & l_a & L'\end{matrix}\right\}\left\{\begin{matrix}\frac{1}{2} & \frac{1}{2} & S\\\frac{1}{2} & \frac{1}{2} & S'\end{matrix}\right\}\Gamma^{(L^\pi S)'}_{ad;cb}~.
\end{equation}
\subsection{Energy optimization with a semidefinite program}
\subsubsection{Interior point method}
The variational problem for the 2DM can be formulated as a so-called semidefinite program \cite{boyd}, a constrained optimization program where it is demanded that certain matrices, which are functions of the variables being optimized, remain positive semidefinite. In our case, there is  a convex subspace of the matrix space, which is called the feasible region, where $\Gamma$, $\mathcal{Q}(\Gamma)$ and $\mathcal{G}(\Gamma)$ are positive semidefinite. In $\Gamma$-space, the direction of energy decrease is given by $\left(-H^{(2)}\right)$ [see Eq. (\ref{tp_ham})]. If the energy is to be minimized, the objective is to go as far as possible in this direction, without leaving the feasible region. The optimized density matrix is on the edge of the feasible region. Computationally, this problem is solved with an interior point method by optimizing the following cost function
\begin{equation}
\label{cost function}
\Phi(\Gamma,t) = \mathrm{Tr}~\Gamma H^{(2)} - t\ln{\det{\mathcal{P}(\Gamma)}}+ C~,
\end{equation}
with
\begin{equation}
\mathcal{P}(\Gamma) = \left(\begin{matrix}\Gamma & 0 & 0\\0 & \mathcal{Q}(\Gamma) & 0 \\ 0 & 0 & \mathcal{G}(\Gamma)\end{matrix}\right)~.
\end{equation}
The constant $C$ has no influence on the solution but is added here in order to take into account the possibility that the matrices have certain explicit zero eigenvalues connected with imposing spin constraints (see the discussion following Eq. (\ref{50}); in this case $C$ can be considered an infinite constant).
Starting from a large value of $t$, (e.g. $t = 1$), the cost function is minimized, and the resulting density matrix is used as a seed vector for the next minimization program with a smaller value of $t$. This procedure continues until convergence is reached for $t \rightarrow 0$, when the density matrix is at the edge of the feasible region. 
\subsubsection{Implementation}
In addition to the positive semidefinite constraints, there are a number of linear constraints which the density matrix has to fulfill (e.g. particle number). These conditions are imposed by direct substitution. Suppose there are a number of linear constraints of the form,
\begin{equation}
\mathrm{Tr}~\Gamma K^{(i)} = k^{(i)}~.
\end{equation}
The way to impose these conditions is to limit the variations to the subspace orthogonal to the $K^{(i)}$'s. Suppose the set of symmetric tp matrices $\{f^i\}$ is an orthogonal basis of this subspace,
\begin{equation}
\mathrm{Tr}~f^if^j = \delta_{ij}\qquad \mathrm{Tr}~f^iK^{(j)} = 0~.
\end{equation}
The tp density matrix can be expanded in the basis,
\begin{equation}
\Gamma = \sum_i \Gamma_i f^i + \mathcal{C}~,
\end{equation}
where $\mathcal{C}$ is a constant matrix obeying the inhomogeneous conditions
\begin{equation}
\mathrm{Tr}~\mathcal{C}K^{(i)} = k^{(i)}~.
\end{equation}
For the minimization of the cost function at a certain value of $t$, Newton's method is used. At a given point $\Gamma_0$ in matrix space, the gradient of the cost function is
\begin{eqnarray}
\frac{\partial \Phi}{\partial \Gamma_i} &=& \mathrm{Tr}~f^iH^{(2)} - t\left[\mathrm{Tr}~f^i \left\{\Gamma_0^{-1} + \mathcal{Q}\left(\mathcal{Q}(\Gamma_0)^{-1}\right) + \mathcal{G}\left(\mathcal{G}(\Gamma_0)^{-1}\right)\right\}\right].
\end{eqnarray}
Using the Hermiticity of the $\mathcal{Q}$ and $\mathcal{G}$ mappings [e.g. $\mathrm{Tr}~\mathcal{Q}(\Gamma)A = \mathrm{Tr}~\mathcal{Q}(A)\Gamma$], the gradient in matrix form reads:
\begin{equation}
\nabla \Phi = \sum_i \frac{\partial \Phi}{\partial \Gamma_i} f^i = \hat{P}\left[H^{(2)} - t\left(\Gamma_0^{-1} + \mathcal{Q}\left(\mathcal{Q}(\Gamma_0)^{-1}\right) + \hat{A}\mathcal{G}\left(\mathcal{G}(\Gamma_0)^{-1}\right)\right)\right]~,
\end{equation}
where $\hat{P}$ is the operator that projects onto the space spanned by the $f^i$'s and $\hat{A}$ is the antisymmetrizer that projects ph space on tp space. The Hessian at $\Gamma_0$ can be written as
\begin{equation}
\mathcal{H}^{ij}=\frac{\partial^2\Phi}{\partial\Gamma_i\partial\Gamma_j} = t\left\{\mathrm{Tr}~\left[f^i\Gamma_0^{-1}f^j\Gamma_0^{-1}\right] + \mathrm{Tr}~\left[\mathcal{Q}(f^i)\mathcal{Q}(\Gamma_0)^{-1}\mathcal{Q}(f^j)\mathcal{Q}(\Gamma_0)^{-1}\right] + \mathrm{Tr}~\left[\mathcal{G}(f^i)\mathcal{G}(\Gamma_0)^{-1}\mathcal{G}(f^j)\mathcal{G}(\Gamma_0)^{-1}\right]\right\}~.
\end{equation}
In Newton's method the search direction $\epsilon$ is found by solving the linear system 
\begin{equation}
\sum_j\frac{\partial^2\Phi}{\partial\Gamma_i\partial\Gamma_j} \epsilon_j = -\frac{\partial \Phi}{\partial \Gamma_i}~.
\end{equation}

This system is solved using the linear conjugate gradient method \cite{golub}. In this method, only one matrix-vector multiplication is needed per iteration. The special structure of the Hessian can be exploited to construct a fast matrix-vector multiplication. The action of the Hessian on a tp matrix $\epsilon$ is
\begin{equation}
\sum_j\mathcal{H}^{ij}\epsilon_j = t\left\{\mathrm{Tr}~\left[f^i\Gamma_0^{-1}\epsilon\Gamma_0^{-1}\right] + \mathrm{Tr}~\left[\mathcal{Q}(f^i)\mathcal{Q}(\Gamma_0)^{-1}\mathcal{Q}(\epsilon)\mathcal{Q}(\Gamma_0)^{-1}\right] + \mathrm{Tr}~\left[\mathcal{G}(f^i)\mathcal{G}(\Gamma_0)^{-1}\mathcal{G}(\epsilon)\mathcal{G}(\Gamma_0)^{-1}\right]\right\}~,
\end{equation}
which can be written in matrix form as,
\begin{equation}
\mathcal{H}\epsilon = t\hat{P}\left[\Gamma_0^{-1}\epsilon\Gamma_0^{-1} + \mathcal{Q}\left(\mathcal{Q}(\Gamma_0)^{-1}\mathcal{Q}(\epsilon)\mathcal{Q}(\Gamma_0)^{-1}\right) + \hat{A}\mathcal{G}\left(\mathcal{G}(\Gamma_0)^{-1}\mathcal{G}(\epsilon)\mathcal{G}(\Gamma_0)^{-1}\right)\right]~.
\end{equation}
It is clear that each conjugate gradient step can be calculated using only manipulations in the tp and ph matrix space.

After the convergence of the conjugate gradient cycle, the direction of the Newton-Raphson step $\epsilon$ is known. A line search in this direction is then performed in order to obtain the minimum of the cost function. Note that one always stays in the feasible region since the cost function goes to $+\infty$ at the edge. 
\subsubsection{Imposing the spin constraints for $\Sigma = 0$}
The spin coupled form of the $\hat{\Sigma}_z$ operator can be written as:
\begin{equation}
\label{S_Z}
\hat{\Sigma}_z = \frac{1}{\sqrt{2}}\sum_a\left[a^\dagger_{a}\otimes\tilde{a}_{a}\right]^1_0~.
\end{equation}
This operator lives in ph-space, and we can force the vector
\begin{equation}
\{\hat{\Sigma}_z\}^S_{ab} = \frac{1}{\sqrt{2}}\delta_{S1}\delta_{ab}~,
\end{equation}
to be an eigenvector of $\mathcal{G}(\Gamma)$ with eigenvalue zero. In doing this we automatically impose the same constraints on $\mathcal{G}(\Gamma)$ for $\Sigma_x$ and $\Sigma_y$ due to the threefold degeneracy of the $S=1$ block of the 2DM. It can be easily seen that in this case the expectation value of the total spin is zero,
\begin{equation}
\bra{\Psi^N}\hat{\Sigma}^2\ket{\Psi^N} = \bra{\Psi^N}\hat{\Sigma}_x^2+\hat{\Sigma}_y^2+\hat{\Sigma}_z^2\ket{\Psi^N} =0~.
\end{equation}
So the condition to be imposed on the density matrix becomes:
\begin{equation}
\label{50}
\sum_S [S]^2\left[\frac{1}{2}\frac{1}{N-1} - (-1)^S\left\{\begin{matrix}\frac{1}{2} & \frac{1}{2} & 1 \\ \frac{1}{2} & \frac{1}{2} & S\end{matrix}\right\}\right]\sum_b\Gamma^S_{ab;cb} = 0~.
\end{equation}
For the projection of a tp density matrix on a spin singlet state, there are as many constraint matrices as there are sp matrix dimensions.
Because of the zero eigenvalues in the $G$-matrix the projected density matrix is on the edge of the feasible region during the whole of the minimization process, and as a result, the cost function is infinity. This can be circumvented by taking the pseudo-inverse of the $G$-matrix, which excludes the $\Sigma_z$-state from the inversion process. This will not alter the result of the program because the contribution of this state to the cost function is constant.
\subsubsection{Imposing the spin constraints for $\Sigma\neq 0$}
For higher-spin multiplets we use the spin-averaged ensemble (see Sec.~\ref{spinensemble}), in which the 2DM has the same simple structure as for the singlet case.
The expectation value of the $\hat{\Sigma}^2$ spin operator is forced to be exact, using the linear constraint
\begin{equation}
\mathrm{Tr}~\Gamma \{\hat{\Sigma}^2\} = \Sigma(\Sigma+1)~,
\end{equation}
where  $\{\hat{\Sigma}^2\}$ is the tp matrix representation of the $\hat{\Sigma}^2$ operator,
\begin{equation}
\{\hat{\Sigma}^2\}^{S}_{ab;cd} = \left[\frac{3}{2}\frac{2-N}{N-1} + S(S+1)\right]
\left(\delta_{ac}\delta_{bd} + (-1)^S\delta_{ad}\delta_{bc}\right)~.
\end{equation}
There is only one linear constraint for nonzero spin, in contrast to the numerous constraints for the projection onto a singlet state.
It can therefore be expected that the spin constraints (i.e.\ the constraints on the 2DM ensuring that it is derivable from a wave function with good total spin)
are less accurate than those for the singlet case. It is, in fact, known how to cure this situation \cite{maz_spin} by considering not the
spin-averaged ensemble but rather the 2DM derived from the highest-weight member ($\mu=\Sigma$) of the multiplet.
Similar to the spin singlet projection, one can then impose the condition that, since the spin-raising ladder operator $\hat{\Sigma}_+$ destroys the
wave function, the $\mathcal{G}(\Gamma)$ matrix must have a zero eigenvalue (with an eigenvector in ph space corresponding to the $\hat{\Sigma}_+$ operator).
In such a highest-weight schemes, the spin restrictions for the $\Sigma\neq 0$ case are put  on the same footing as for the singlet case; in fact, the highest-weight and the spin-averaged ensemble scheme are equivalent for the singlet case.
However, the highest-weight scheme for $\Sigma\neq 0$ requires one to keep track of more matrices and is computationally more demanding by about a factor of 10. We therefore used the ensemble scheme even for the nonsinglets (i.e.\ the Si atom), though we checked some cases with the highest-weight method for the spin.
\subsubsection{Spin and angular momentum projection}
When angular momentum is taken into account, everything becomes a bit more complicated, but the principles are the same as in the last paragraph. It can be shown that in a spin-and-angular-momentum-coupled basis the $z$-projections of $\Sigma$ and $\Lambda$ become
\begin{eqnarray}
\Sigma_z &=& \frac{1}{\sqrt{2}}\sum_{nl}[l]\left[a_{nl}^\dagger\otimes\tilde{a}_{nl}\right]^{\left(0^+1\right)}~,\\
\Lambda_z &=& \sqrt{\frac{2}{3}}\sum_{nl}[l]\hat{l}\left[a_{nl}^\dagger\otimes\tilde{a}_{nl}\right]^{\left(1^+0\right)}~.
\end{eqnarray}
Following the same argument as before, it can be imposed that the density matrix is derivable from an eigenstate with zero eigenvalue of respectively the $\Sigma$ and $\Lambda$ operators when 
\begin{eqnarray}
\sum_{c}[l_c]\mathcal{G}\left(\Gamma\right)^{\left(0^+1\right)}_{ab;cc} &=& 0~,\\
\sum_{c}[l_c]\hat{l}_c\mathcal{G}\left(\Gamma\right)^{\left(1^+0\right)}_{ab;cc} &=& 0~.
\end{eqnarray}
This can be translated into linear constraints on the 2DM, which are given in the Appendix. 
The projection on spin and angular momentum not equal to zero is again a less strict condition. The expectation values of $\Lambda$ and $\Sigma$ are projected on the desired values:
\begin{eqnarray}
\mathrm{Tr}~\Gamma \{\hat{\Sigma}^2\} &=& \Sigma\left(\Sigma + 1\right)~,\\
\mathrm{Tr}~\Gamma \{\hat{\Lambda}^2\} &=&  \Lambda\left(\Lambda + 1\right)~,
\end{eqnarray}
where the $\{\hat{\Sigma}^2\}$ and $\{\hat{\Lambda}^2\}$ are the tp matrix representations of the $\hat{\Sigma^2}$ and $\hat{\Lambda^2}$ operators respectively.
\begin{eqnarray}
\{\hat{\Sigma}^2\}^{(L^\pi S)}_{ab;cd} &=& \left[\frac{3}{2}\frac{2-N}{N-1} + \hat{S}^2\right]\\
\nonumber&&\times\left(\delta_{ac}\delta_{bd} + (-1)^{L+S+l_a+l_b}\delta_{ad}\delta_{bc}\right)~,\\
\{\hat{\Lambda}^2\}^{(L^\pi S)}_{ab;cd} &=& \left[\frac{2-N}{N-1}\left(\hat{l}_a^2 + \hat{l}_b^2\right) + \hat{L}^2\right]\\
\nonumber&&\times\left(\delta_{ac}\delta_{bd} + (-1)^{L+S+l_a+l_b}\delta_{ad}\delta_{bc}\right)~.
\end{eqnarray}
\section{Results and discussion}
Using the method explained in the previous section, the isoelectronic series of Be, Ne and Si were calculated from the neutral atom up to a central charge $Z = 28$. Beryllium and neon are both elements with a singlet ground state. In the silicon ground state the total spin and angular momentum are both one, which allows us to assess the quality of the spin and angular momentum constraints for ${\Sigma,\Lambda} \neq 0$. In order to study the basis set dependence, the properties of the ground state of the Be and Ne series were calculated in a cc-pVDZ, a cc-pVTZ and a cc-pVQZ basis set \cite{dunning}. The Si series was  only calculated in a cc-pVDZ and a cc-pVTZ basis set \cite{woon}. We used spherical harmonic (and not Cartesian) basis functions throughout. With the density matrices obtained from the SDP, several properties were studied. These are compared to estimates for non-relativistic energies based on experimental data \cite{davidson,chakravorty}, and to the results of coupled cluster (CCSD) calculations, and in some cases, with full-configuration-interaction (CI) calculations. 

The basis functions used were those of the neutral atom, but with a rescaling $r\rightarrow rZ/N$ for the positive ions with $Z > N$.
The CCSD and full-CI results were obtained using the MOLPRO program \cite{molpro}.
\subsection{Ground-state energy}
The ground-state energies, calculated with various basis sets and methods, are shown in 
Tables \ref{gse_Be}, \ref{gse_Ne} and \ref{gse_Si} for the Be, Ne, and Si isoelectronic series, respectively.    
Even in the best case (Be in cc-pVQZ), the calculated energies are at least 20 mhartree removed from the experimental estimate in \cite{davidson,chakravorty}. 
This is due to the difficulty of describing the interelectronic cusp in the exact wave function using finite sp basis sets.
\begin{sidewaystable*}
\caption{\label{gse_Be} The ground-state energies of the Be series in the cc-pV(DTQ)Z basis sets using different methods.}
\begin{ruledtabular}
\begin{tabular}{cccccccccccccc}
Z&\multicolumn{4}{c}{cc-pVDZ}&\multicolumn{4}{c}{cc-pVTZ}&\multicolumn{4}{c}{cc-pVQZ basis}&expt.\\
&SDP&HF&CCSD&full-CI&SDP&HF&CCSD&full-CI&SDP&HF&CCSD&full-CI&\\
\hline
4&-14.617473&-14.572338&-14.617369&-14.61741&-14.625431&-14.572873&-14.623559&-14.62381&-14.642807&-14.572968&-14.639589&-14.640124&-14.66736\\
5&-24.275712&-24.216056&-24.27566&-24.275684&-24.300695&-24.234557&-24.299207&-24.29943&-24.321254&-24.236385&-24.317643&-24.31822&-24.34892\\
6&-36.387458&-36.316267&-36.387421&-36.387439&-36.473162&-36.394215&-36.471944&-36.47214&-36.500934&-36.40257&-36.497178&-36.497761&-36.53493\\
7&-50.940925&-50.860695&-50.940896&-50.940909&-51.137349&-51.045734&-51.136311&-51.136486&-51.177145&-51.065945&-51.173335&-51.173918&-51.22284\\
8&-67.931909&-67.844323&-67.931884&-67.931896&-68.290965&-68.186797&-68.290046&-68.290206&-68.347448&-68.22364&-68.343621&-68.344203&-68.41171\\
9&-87.358767&-87.265015&-87.358746&-87.358755&-87.932793&-87.816285&-87.931958&-87.932107&-88.010503&-87.87405&-88.00667&-88.007254&-88.10113\\
10&-109.22078&-109.12175&-109.22076&-109.22077&-110.06209&-109.93353&-110.06132&-110.06146&-110.16555&-110.01625&-110.16171&-110.1623&-110.29089\\
11&-133.51761&-133.414&-133.51759&-133.5176&-134.67837&-134.53811&-134.67764&-134.67778&-134.81213&-134.64967&-134.8083&-134.80889&-134.98088\\
12&-160.24908&-160.14145&-160.24906&-160.24907&-161.7813&-161.62969&-161.78061&-161.78074&-161.95&-161.77398&-161.94616&-161.94677&-162.17102\\
13&-189.41511&-189.30392&-189.41509&-189.4151&-191.37066&-191.20808&-191.37&-191.37011&-191.57899&-191.38895&-191.57514&-191.57575&-191.86127\\
14&-221.01564&-220.90129&-221.01563&-221.01564&-223.44627&-223.27309&-223.44563&-223.44574&-223.699&-223.49443&-223.69514&-223.69577&-224.0516\\
15&-255.05067&-254.93347&-255.05066&-255.05066&-258.00801&-257.82461&-258.0074&-258.00751&-258.30998&-258.09031&-258.3061&-258.30675&-258.742\\
16&-291.52018&-291.4004&-291.52017&-291.52017&-295.05582&-294.86255&-295.05522&-295.05533&-295.4119&-295.17653&-295.40801&-295.40867&-295.93244\\
17&-330.42417&-330.30206&-330.42416&-330.42416&-334.58961&-334.38682&-334.58904&-334.58913&-335.00476&-334.75303&-335.00085&-335.00153&-335.62293\\
18&-371.76264&-371.63841&-371.76263&-371.76263&-376.60935&-376.39737&-376.60879&-376.60888&-377.08857&-376.81977&-377.08463&-377.08533&-377.81344\\
19&-415.5356&-415.40942&-415.53559&-415.53559&-421.115&-420.89414&-421.11445&-421.11454&-421.66334&-421.37673&-421.65938&-421.66011&-422.50398\\
20&-461.74304&-461.61508&-461.74303&-461.74304&-468.10654&-467.87712&-468.106&-468.10609&-468.72912&-468.42388&-468.72513&-468.72588&-469.69455\\
21&-510.38498&-510.25537&-510.38498&-510.38498&-517.58394&-517.34625&-517.58342&-517.5835&-518.28593&-517.96121&-518.28191&-518.28269&-519.38513\\
22&-561.46143&-561.3303&-561.46142&-561.46142&-569.5472&-569.30152&-569.54669&-569.54677&-570.33383&-569.9887&-570.32977&-570.33058&-571.57572\\
23&-614.97237&-614.83983&-614.97237&-614.97237&-623.99631&-623.7429&-623.99581&-623.99589&-624.87286&-624.50634&-624.86877&-624.86961&-626.26633\\
24&-670.91783&-670.78398&-670.91783&-670.91783&-680.93126&-680.67038&-680.93077&-680.93084&-681.90309&-681.51414&-681.89895&-681.89983&-683.45695\\
25&-729.29781&-729.16274&-729.2978&-729.2978&-740.35204&-740.08394&-740.35156&-740.35163&-741.42459&-741.01206&-741.4204&-741.42132&-743.14758\\
26&-790.1123&-789.97609&-790.11229&-790.1123&-802.25866&-801.98356&-802.25818&-802.25825&-803.43742&-803.00013&-803.43318&-803.43414&-805.33822\\
27&-853.36132&-853.22404&-853.36131&-853.36131&-866.65111&-866.36925&-866.65064&-866.65071&-867.94167&-867.47832&-867.93738&-867.93838&-870.02886\\
28&-919.04486&-918.90659&-919.04486&-919.04486&-933.5294&-933.24098&-933.52894&-933.529&-934.93744&-934.44663&-934.93309&-934.93413&-937.21951\\
\end{tabular}
\end{ruledtabular}
\end{sidewaystable*}
\begin{sidewaystable*}
\caption{\label{gse_Ne} The ground-state energies of the Ne series in the cc-pV(DTQ)Z basis sets using different methods.}
\begin{ruledtabular}
\begin{tabular}{cccccccccccc}
Z&\multicolumn{4}{c}{cc-pVDZ}&\multicolumn{3}{c}{cc-pVTZ}&\multicolumn{3}{c}{cc-pVQZ basis}&expt.\\
&SDP&HF&CCSD&full-CI&SDP&HF&CCSD&SDP&HF&CCSD&\\
\hline
10&-128.70843&-128.48878&-128.67964&-128.68088&-128.86088&-128.53186&-128.81081&-128.92686&-128.54347&-128.87106&-128.9376\\
11&-161.80049&-161.59591&-161.77283&-161.77411&-161.97703&-161.65496&-161.92829&-162.05038&-161.67155&-161.99595&-162.0659\\
12&-198.88784&-198.70208&-198.86199&-198.86309&-199.11372&-198.79861&-199.06598&-199.19913&-198.82303&-199.14502&-199.2204\\
13&-239.97194&-239.80393&-239.94802&-239.94883&-240.26728&-239.9582&-240.22028&-240.36392&-239.98965&-240.30977&-240.3914\\
14&-285.04223&-284.88894&-285.02004&-285.02061&-285.43166&-285.12786&-285.38525&-285.53886&-285.16605&-285.48453&-285.5738\\
15&-334.08381&-333.94195&-334.06299&-334.06338&-334.6021&-334.30313&-334.55624&-334.72067&-334.3492&-334.66615&-334.7642\\
16&-387.08194&-386.94882&-387.06219&-387.06246&-387.77553&-387.48107&-387.73017&-387.90757&-387.53731&-387.85281&-387.9608\\
17&-444.02427&-443.89781&-444.00531&-444.00551&-444.95002&-444.6598&-444.9051&-445.09826&-444.72921&-445.04331&-445.1622\\
18&-504.90101&-504.77972&-504.88268&-504.88282&-506.12426&-505.83808&-506.07977&-506.29194&-505.92402&-506.23681&-506.3673\\
19&-569.70474&-569.58754&-569.68689&-569.68701&-571.29734&-571.01502&-571.25329&-571.48788&-571.12105&-571.43261&-571.5754\\
20&-638.42981&-638.31595&-638.41239&-638.41248&-640.46864&-640.18995&-640.42496&-640.68561&-640.31973&-640.63013&-640.7891\\
21&-711.07205&-710.96091&-711.05497&-711.05505&-713.63756&-713.36231&-713.59424&-713.88444&-713.51958&-713.8289&-713.9988\\
22&-787.6282&-787.51937&-787.61143&-787.6115&-790.80365&-790.53161&-790.76063&-791.08414&-790.72018&-791.02848&-791.2132\\
23&-868.09589&-867.98895&-868.07932&-868.07938&-871.96637&-871.69743&-871.9237&-872.28418&-871.92117&-872.22853&-872.4291\\
24&-952.47304&-952.36781&-952.45674&-952.45679&-957.12544&-956.85936&-957.08305&-957.48456&-957.12224&-957.42872&-957.6463\\
25&-1040.7584&-1040.6545&-1040.7422&-1040.7422&-1046.2804&-1046.0171&-1046.2383&-1046.6846&-1046.3231&-1046.6288&-1046.8646\\
26&-1132.9505&-1132.8479&-1132.9345&-1132.9345&-1139.431&-1139.1702&-1139.3892&-1139.8845&-1139.5236&-1139.8285&-1140.0838\\
27&-1229.0485&-1228.9471&-1229.0327&-1229.0328&-1236.5769&-1236.3184&-1236.5353&-1237.0837&-1236.7235&-1237.0277&-1237.3039\\
28&-1329.0518&-1328.9514&-1329.0361&-1329.0361&-1337.7178&-1337.4616&-1337.6764&-1338.2821&-1337.9226&-1338.2261&-1338.5247\\
\end{tabular}
\end{ruledtabular}
\end{sidewaystable*}
\begin{sidewaystable*}
\caption{\label{gse_Si} The ground-state energies of the Si series in the cc-pV(DT)Z basis sets using different methods. The results under SDP were calculated using the ensemble avaraged spin projection, those under SDP$^*$ were calculated using the maximal weight method.}
\begin{ruledtabular}
\begin{tabular}{ccccccccc}
Z&\multicolumn{4}{c}{cc-pVDZ}&\multicolumn{3}{c}{cc-pVTZ}&expt.\\
&SDP&SDP$^*$&HF&CCSD&SDP&HF&CCSD&\\
\hline
14&-288.93962&-288.92921&-288.84644&-288.91895&-289.02515&-288.85215&-288.9835&-289.359\\
15&-340.36765&&-340.27338&-340.34709&-340.50472&-340.33467&-340.46205&-340.872\\
16&-396.10801&&-396.01679&-396.08749&-396.43974&-396.27384&-396.39711&-396.869\\
17&-456.09635&&-456.00926&-456.0759&-456.80372&-456.64236&-456.7617&-457.337\\
18&-520.29362&-520.27860&-520.21067&-520.27348&-521.58&-521.42294&-521.53858&-522.269\\
19&-588.68067&&-588.60149&-588.66097&-590.75683&-590.60373&-590.7159&-591.66\\
20&-661.24791&&-661.17202&-661.22871&-664.32396&-664.17433&-664.28328&-665.507\\
21&-737.99017&&-737.91714&-737.97148&-742.2714&-742.12467&-742.23072&-743.808\\
22&-818.90446&-818.88808&-818.83388&-818.88621&-824.58949&-824.44532&-824.54882&-826.559\\
23&-903.98889&&-903.92042&-903.97105&-911.27007&-911.12778&-911.22912&-913.762\\
24&-993.24222&&-993.1756&-993.22475&-1002.3054&-1002.1648&-1002.2643&-1005.413\\
25&-1086.6636&&-1086.5986&-1086.6465&-1097.6895&-1097.5501&-1097.6482&-1101.513\\
26&-1184.2525&-1184.2381&-1184.1889&-1184.2357&-1197.4173&-1197.2789&-1197.3759&-1202.061\\
27&-1286.0084&&-1285.9461&-1285.9919&-1301.4847&-1301.347&-1301.4431&-1307.057\\
28&-1391.9311&&-1391.8699&-1391.9147&-1409.8882&-1409.7512&-1409.8466&-1416.5\\
\end{tabular}
\end{ruledtabular}
\end{sidewaystable*}
\begin{figure}
\includegraphics[scale=0.5]{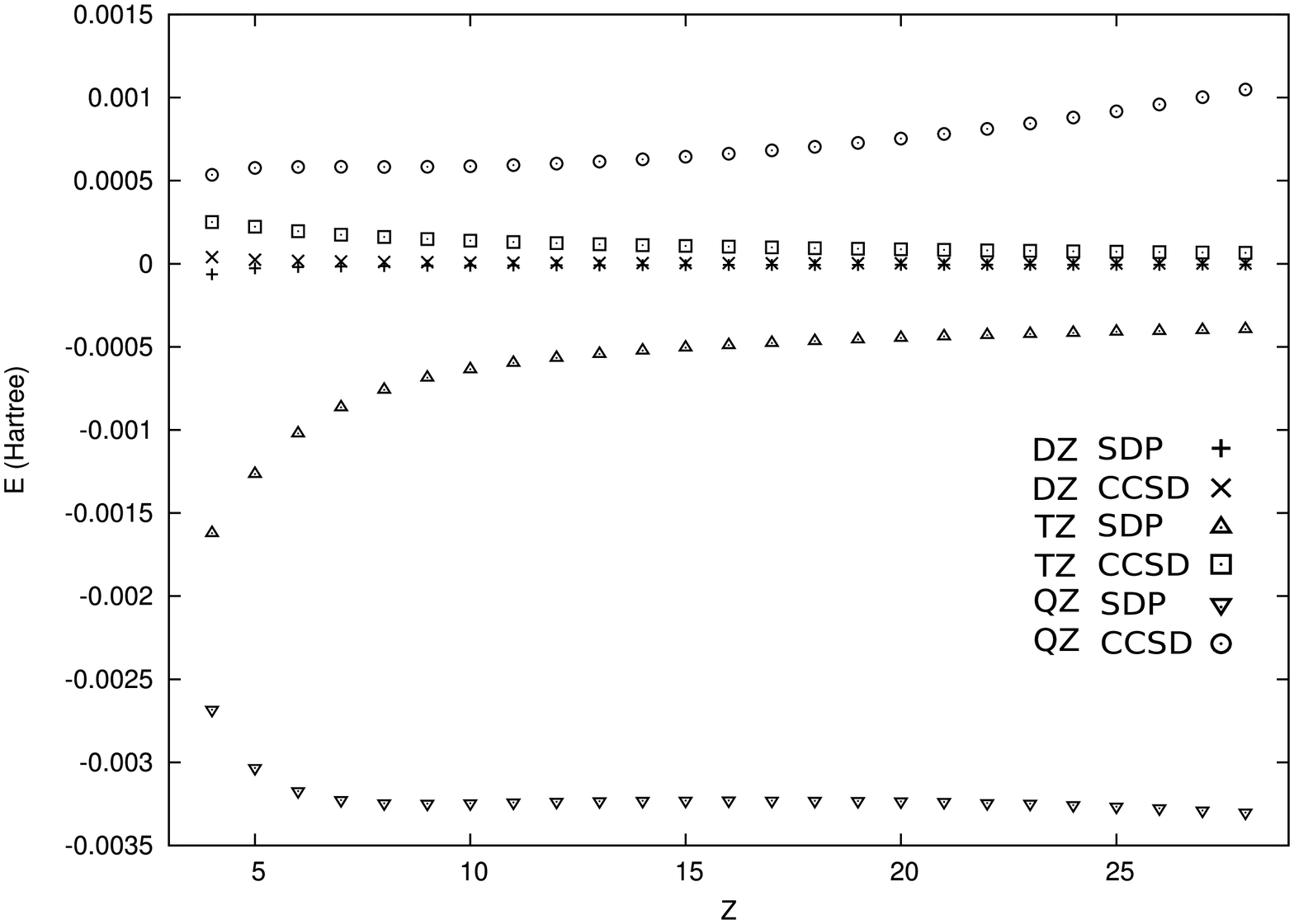}
\caption{\label{beryl_diff_fci} Difference between approximate (CCSD or SDP) and full-CI energies for the Be series in all three basis sets.}
\end{figure}
More relevant is the difference between the SDP (and CCSD) energies as compared to full-CI in the same basis set. 
This is shown in Figure~\ref{beryl_diff_fci} for the case of the Be series. Note that the CCSD energy is always above, the SDP energy below, the full-CI 
energy. For SDP, this simply reflects the nature of the variational problem. For the smallest cc-pVDZ basis set, SDP and CCSD have about the same level of accuracy. 
The difference with full-CI grows as the basis set size increases for both CCSD and SDP, but this effect is worse for the SDP. 

As far as the $Z$-dependence is concerned, the trend differs markedly for the cc-pV(D,T)Z and for the cc-pVQZ basis set.   
As $Z$ increases there is a growing accuracy for the smaller basis sets in both CCSD and SDP, whereas for cc-pVQZ 
the accuracy decreases for CCSD and becomes constant for SDP. The reason for this difference is not clear, though it is probably 
connected to the incipient degeneracy of the $2s$ and $2p$ states and the quality of its description in the various basis sets, as is more fully described in the 
next Section. It should be noted that the SDP results are overall very accurate, even in the worst case ($Z=28$, cc-pVQZ) differing less than 
3 mhartree from full-CI.     

For the Ne series, full-CI calculations were only possible in the cc-pVDZ basis. From the results collected in Table~\ref{gse_Ne} it is seen that the SDP accuracy 
is significantly less than for Be, the largest deviation (28 mhartree) appearing for the neutral atom. This is actually consistent with PQG-condition SDP results 
for molecules, so it is likely that because of the small number of electrons the Be results are not representative. This is also borne out by the Si results in 
Table~\ref{gse_Si}, showing a maximal deviation between CCSD and SDP energies of 21 mhartree for the neutral atom.
\subsection{Correlation energy}
The correlation energies were calculated by taking the difference of the SDP and CCSD energies with the hartree-Fock results in the same basis set. 
The results labeled ``experimental'' are the estimates in \cite{chakravorty}.
\subsubsection{Beryllium series}
\begin{figure}
\includegraphics[scale=0.6]{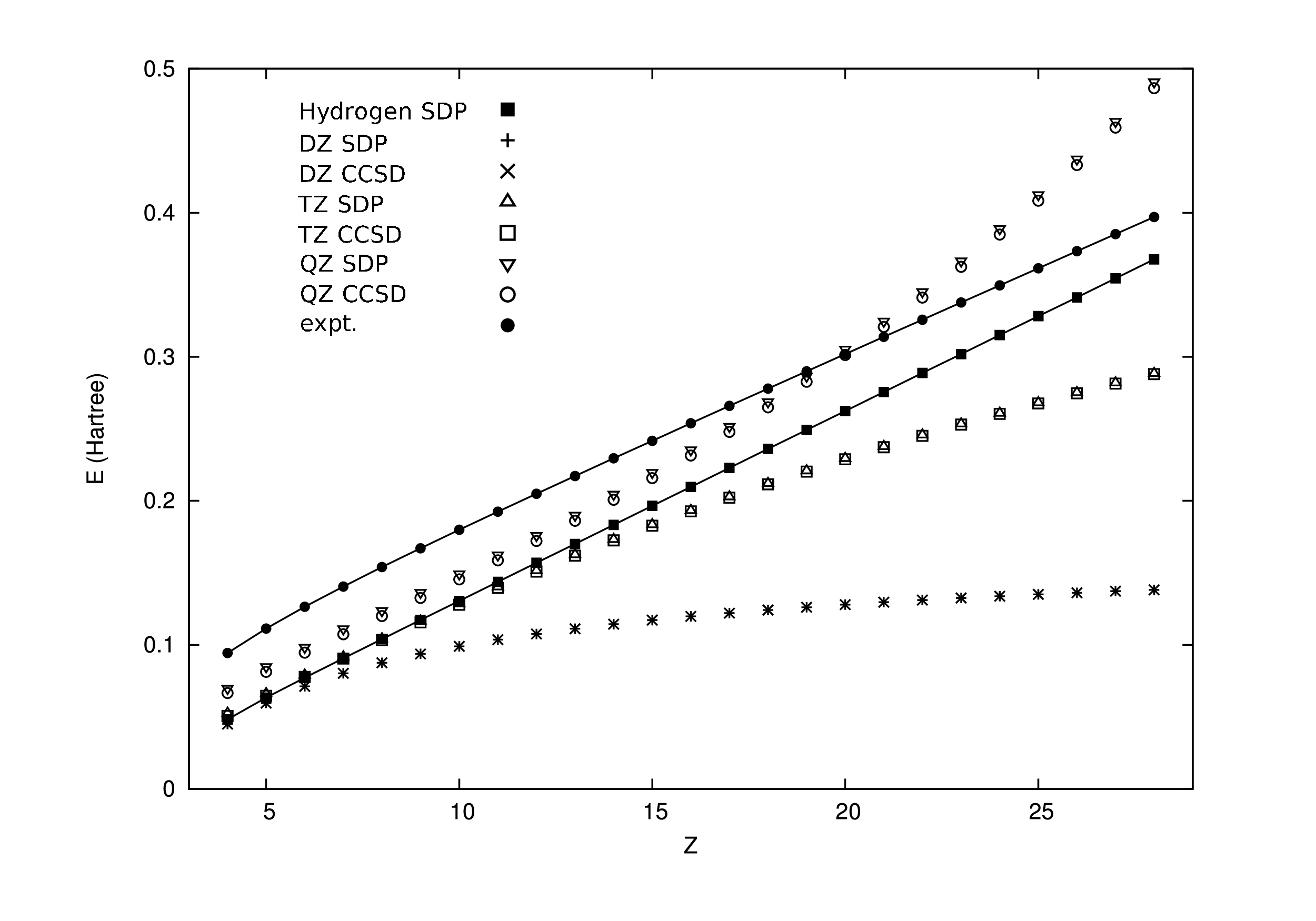}
\caption{\label{beryl_corr_ener} The SDP correlation energy for the Be series in all three basis sets, and in a rescaled basis set that exhibits hydrogen-like behaviour (degeneracy between the $2s$ and $2p$ level). For comparison, the CCSD and experimental values are also shown.}
\end{figure}

In Fig.~\ref{beryl_corr_ener} the SDP correlation energy is shown as a function of central charge $Z$ for the 
different basis sets. Note that on the plot the difference between the CCSD and full-CI correlation energies would not be visible. 
The experimental curve is linear in $Z$, as a direct consequence of the near-degeneracy of the ground state \cite{chakravorty}.
One can calculate a perturbative series expansion of the exact and hartree-Fock energy in powers of $\frac{1}{Z}$; the corresponding series for the correlation 
energy starts with a constant if the hydrogenic ground state is nondegenerate, or with a linear term in $Z$ in case of degeneracy.   
The SDP correlation energy does not follow this trend: it goes linear in the beginning, but becomes concave in the cc-pVDZ and cc-pVTZ basis, or convex in the cc-pVQZ basis. This failure, however, is not related to the SDP method as the trend is the same in full-CI. It simply reflects the fact that the incipient degeneracy is not well described in these basis sets. This can also be seen by calculating the $Z=1$ hydrogen spectrum (corresponding to the 
$Z\rightarrow \infty$ situation, when the electron-electron interaction can be neglected) in the basis sets: the $2s$ and $2p$ energies are not degenerate, but differ by 5.8 mhartree (cc-pVDZ), 2.0 mhartree (cc-pVTZ) and -2.3 mhartree (cc-pVQZ). Note that for CC-pVQZ the $2p$ energy actually drops below the $2s$ energy, explaining the different (convex/concave) behavior of the 
curves. To make sure we also performed calculations in the cc-pVDZ basis after rescaling  ($r\rightarrow\alpha r$) it in such a way that the hydrogenic 
$2s$-$2p$ degeneracy is exact. In this basis the SDP correlation energy (also shown in Fig.~\ref{beryl_corr_ener}) indeed has the correct linear behavior. It is clear from the above discussion that SDP is indeed capable of providing accurate correlation energies in the presence of near degeneracies, when other many-body techniques (like density functional theory or MP2) can fail.
\subsubsection{Neon series}
\begin{figure}
\includegraphics[scale=0.6]{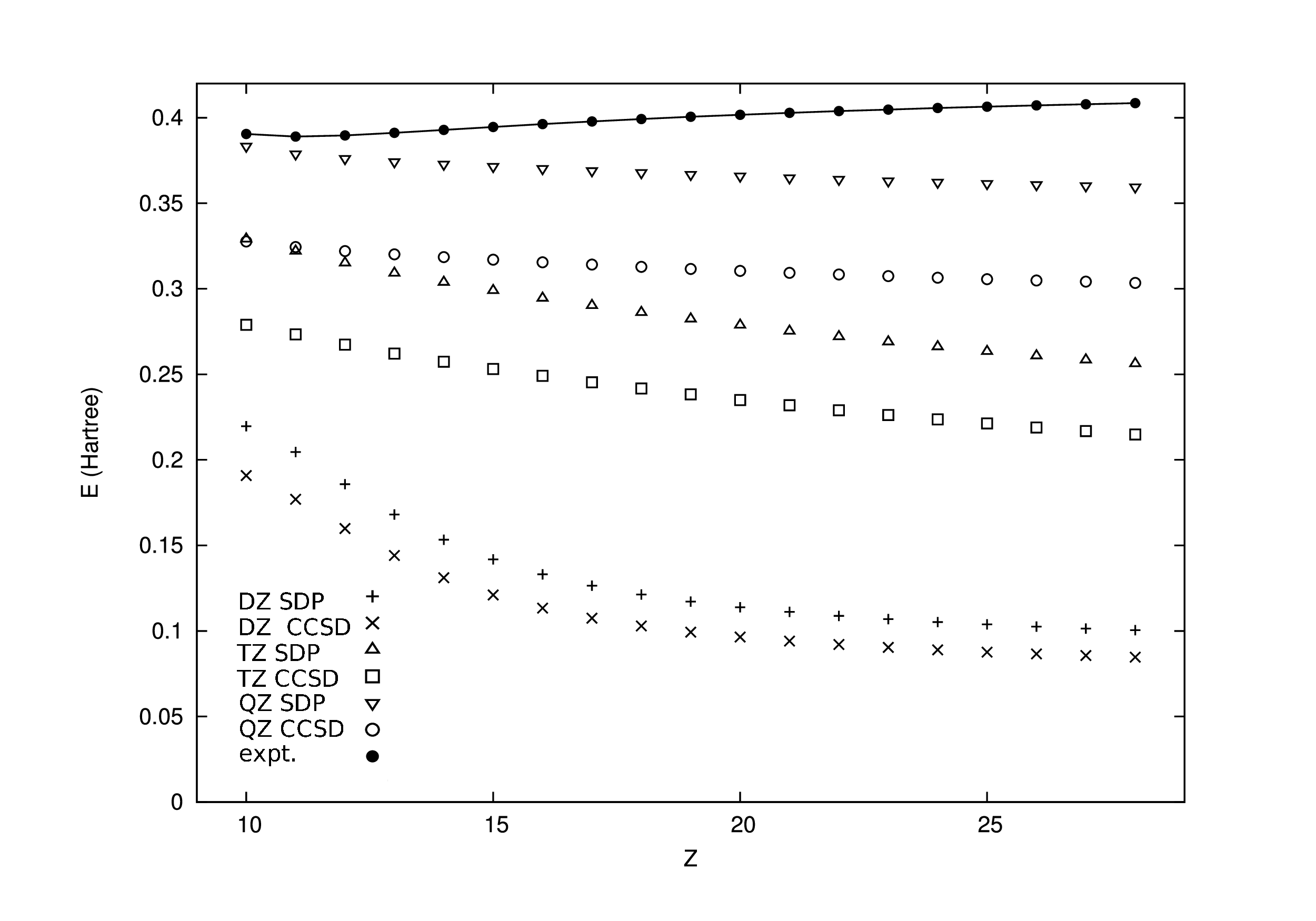}
\caption{\label{neon_corr_ener} The SDP correlation energy for the Ne series in all three basis sets. For comparison, the CCSD and experimental values are also shown.}
\end{figure}

In Fig. \ref{neon_corr_ener} the correlation energy is shown for all three basis sets as a  function of $Z$. Because Ne is a closed shell atom, there is no 
near-degeneracy for large $Z$ values and the exact correlation should be asymptotically constant in $Z$, as is indeed visible in the experimental curve. 
Due to basis set effects, this constant behavior is imperfectly realized, but the SDP follows the same trends as CCSD for all basis sets.  
Note that the approximation to a constant behavior at large $Z$ is best for the largest basis set. The decrease in correlation energy for increasing $Z$, 
in contrast to the slight rise in the experimental correlation energy, can be attributed   
to the fact that the basis sets were optimized for the neutral atom. While the rescaling procedure fixes the nuclear cusp, the resulting basis set is 
obviously far from optimal for highly charged ions.

\subsubsection{Silicon series}

\begin{figure}
\includegraphics[scale=0.6]{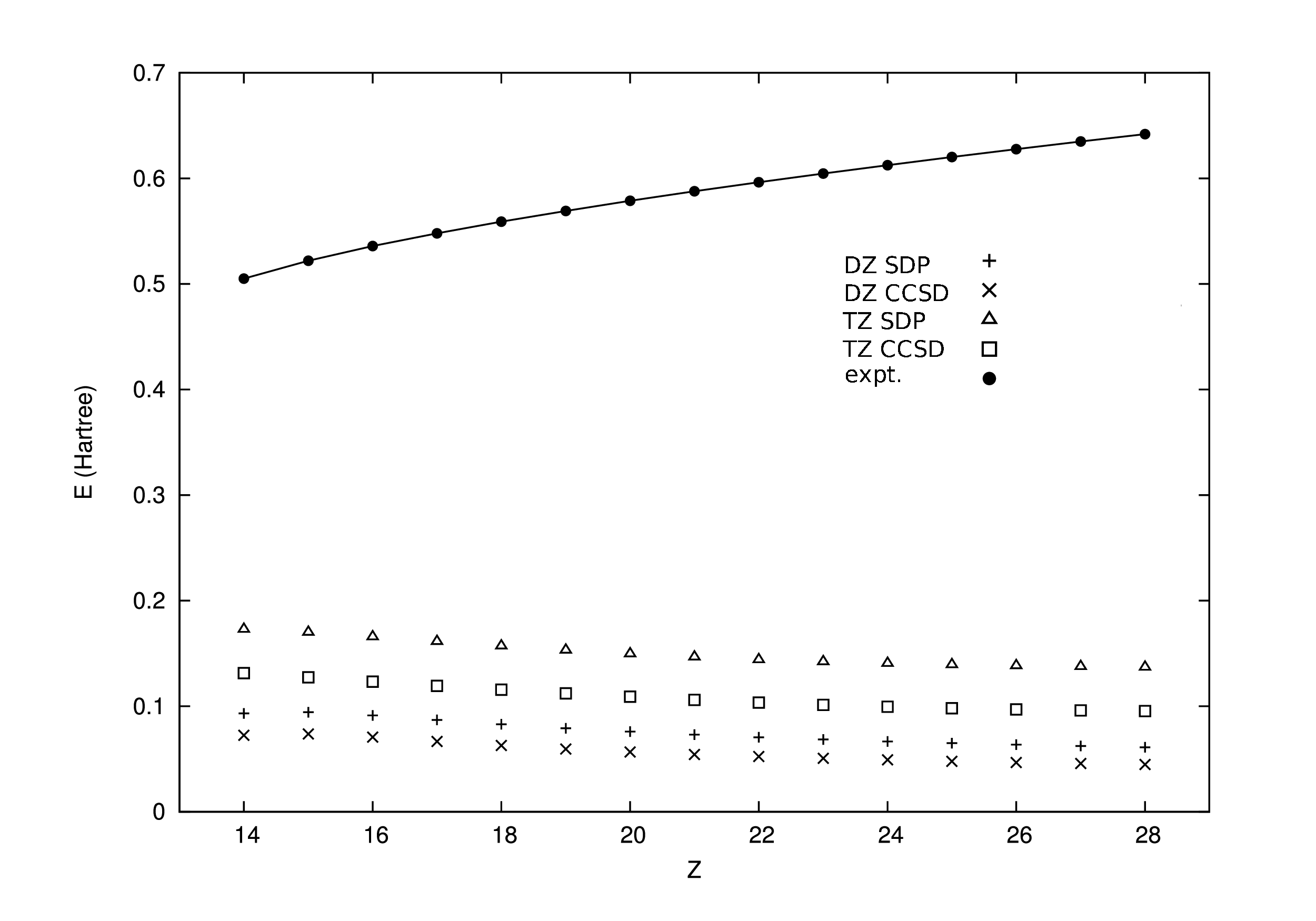}
\caption{\label{Silicon_corr_ener} The SDP correlation energy for the Si series in the cc-pVDZ and cc-pVTZ basis set. For comparison, the CCSD and experimental values are also shown.}
\end{figure}
For silicon, only the cc-pVDZ and cc-pVTZ basis have been used [Fig. \ref{Silicon_corr_ener}]. As was the case for Be, the theoretical linear rise with $Z$ is thwarted by 
imperfections in the basis sets. However the SDP correlation energy closely tracks the CCSD one. The Si ground state is a spin triplet. The results in Table~\ref{gse_Si} have been obtained using the spin-averaged ensemble, as explained in Sec~II. In order to assess the quality of the spin constraints, we have also performed calculations using the highest-weight method, for $Z$=14, 18, 22, and 26 with the cc-pVDZ basis set. The resulting energies are also reported in Table~\ref{gse_Si}. The energy differences between the approaches are sizeable, with differences as large as 20 mhartree,
reflecting the weaker nature of the spin constraints imposed in the spin-averaged scheme.         
However, the discrepancy between the two approaches is stable for increasing $Z$. 
\subsection{Ionization energies}
\begin{figure}
\includegraphics[scale=0.5]{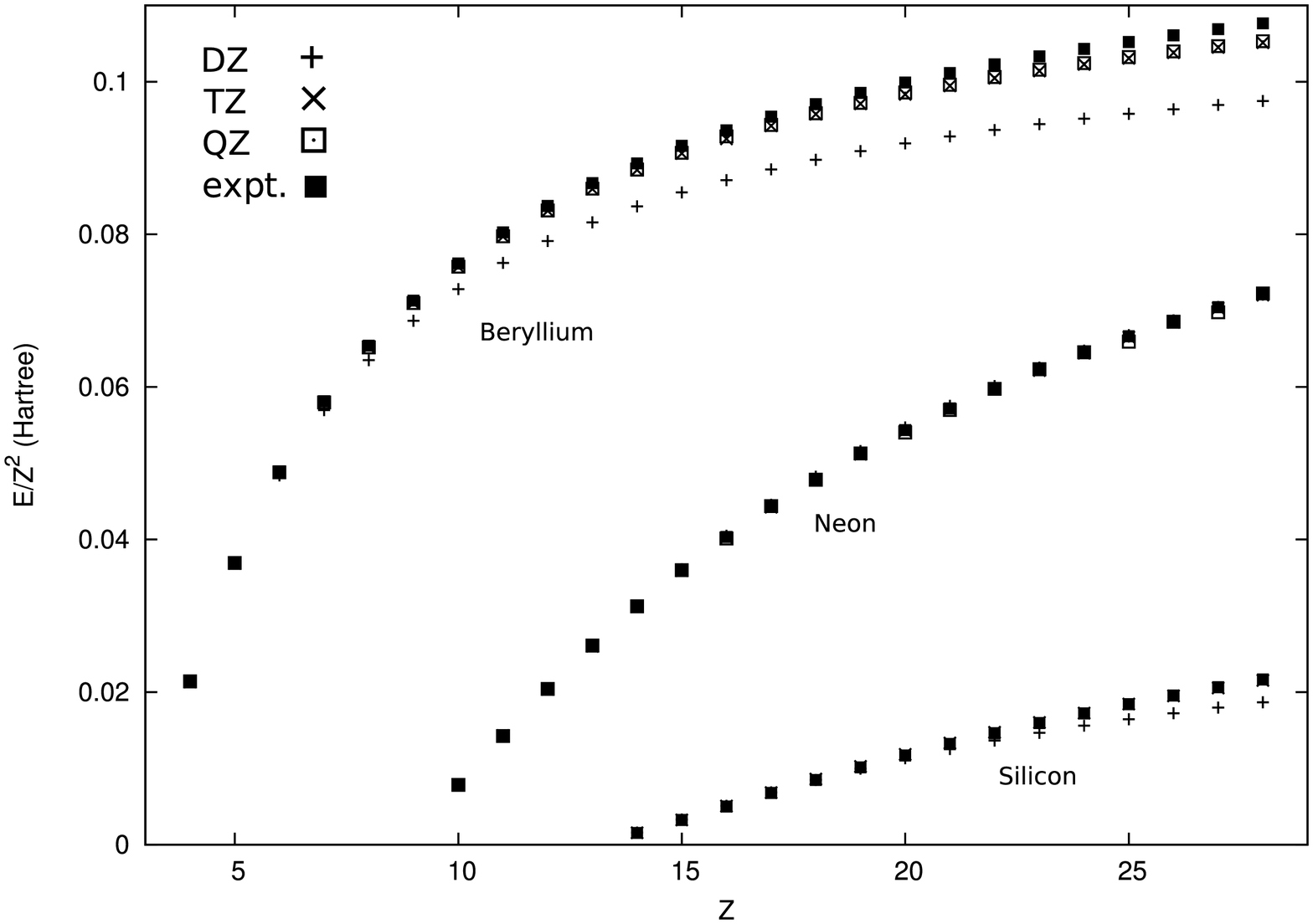}
\caption{\label{ion_E} The ionization energy scaled with $\frac{1}{Z^2}$, for the Be, Ne and Si series in the different basis sets compared with experimental results.}
\end{figure}
Other properties can be used to gauge the quality of the 2DM, e.g., the  ionization energies of the different atomic ions, which can be easily calculated using the extended Koopmans' theorem (EKT) \cite{day,smith,morrell}. The EKT provides a single particle picture of the ground state, with sp energies and spectroscopic factors. The ionization energies are shown in Figure~\ref{ion_E}; the agreement between calculated and  
experimental values is very good, pointing to the realistic nature of the variationally obtained 2DM. 
The good agreement with experiment reflects the fact that the error in the description of the interelectronic cusp largely cancels since the ionization energy is 
an energy difference. For Be and Ne it is clear that the basis set limit is nearly reached at the cc-pVTZ -- cc-pVQZ level. Even for Si the experimental ionization energy 
is closely reproduced.
\subsection{Correlated hartree-Fock-like single-particle energies}
\begin{figure}
\includegraphics[scale=0.5]{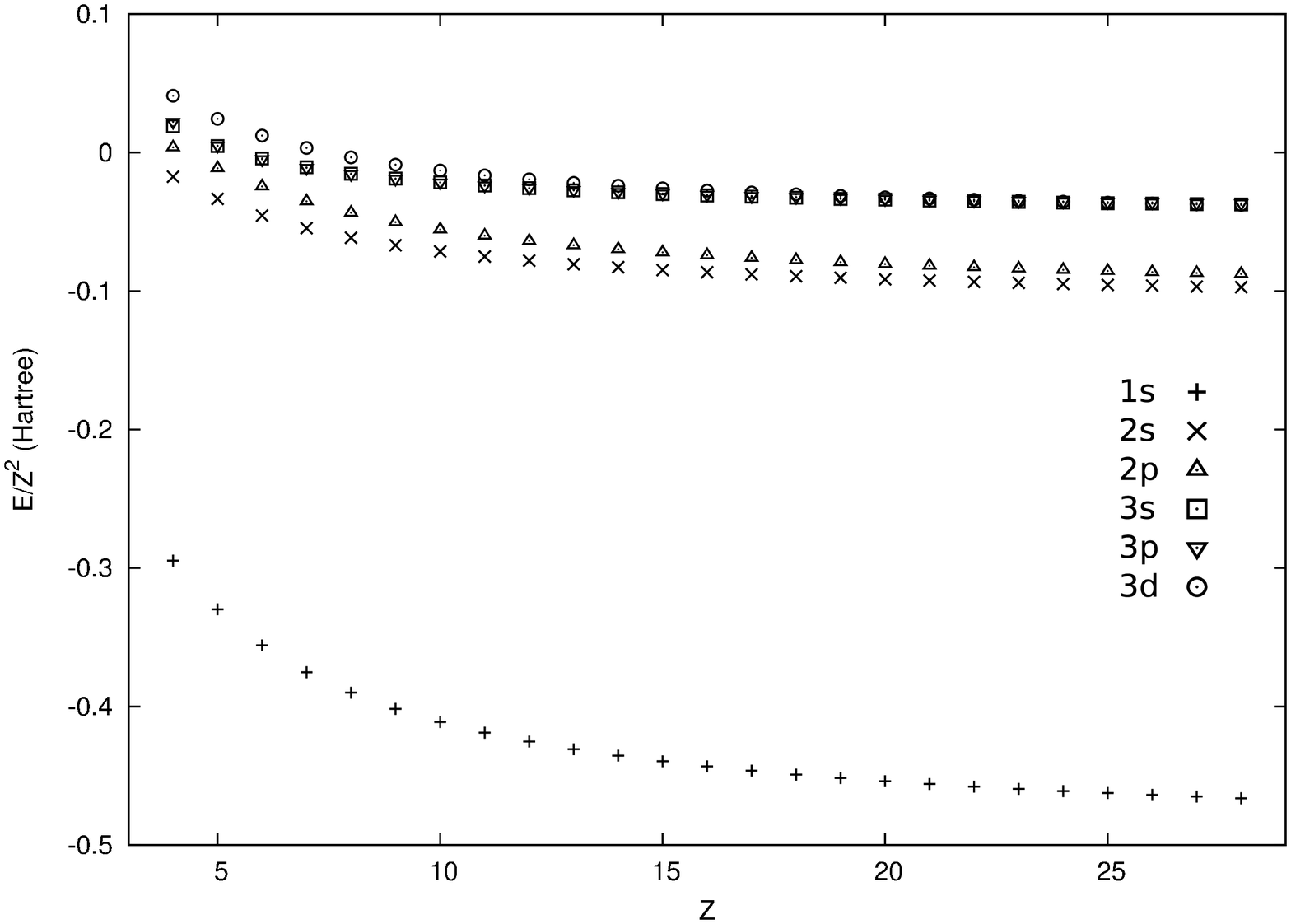}
\caption{\label{sp_levels} The single particle levels obtained in a correlated hartree-Fock-like scheme (see text) for the Be series in a cc-pVDZ basis set.}
\end{figure}
A different  sp picture is given by the correlated hartree-Fock-like sp orbitals and energies. These are constructed by diagonalizing the sp Hamiltonian:
\begin{equation}
h_{\alpha\gamma} = \left(T + U\right)_{\alpha\gamma} + \sum_{\beta\delta}V_{\alpha\beta;\gamma\delta}\rho_{\beta\delta}~,
\end{equation}
where the first-order density matrix (1DM) ($\rho$) is constructed from the variationally determined 2DM. 
As an example of this method, the sp energies for the isoelectronic series of Be in a cc-pVDZ basis are shown in 
Figure~\ref{sp_levels}. Notice that when $Z$ increases, the energy levels approach those of the hydrogen atom. 
Similar behavior is present for the other basis sets and for the Ne and Si isoelectronic series. 
\subsection{Natural occupations}
\begin{figure}
\includegraphics[scale=0.5]{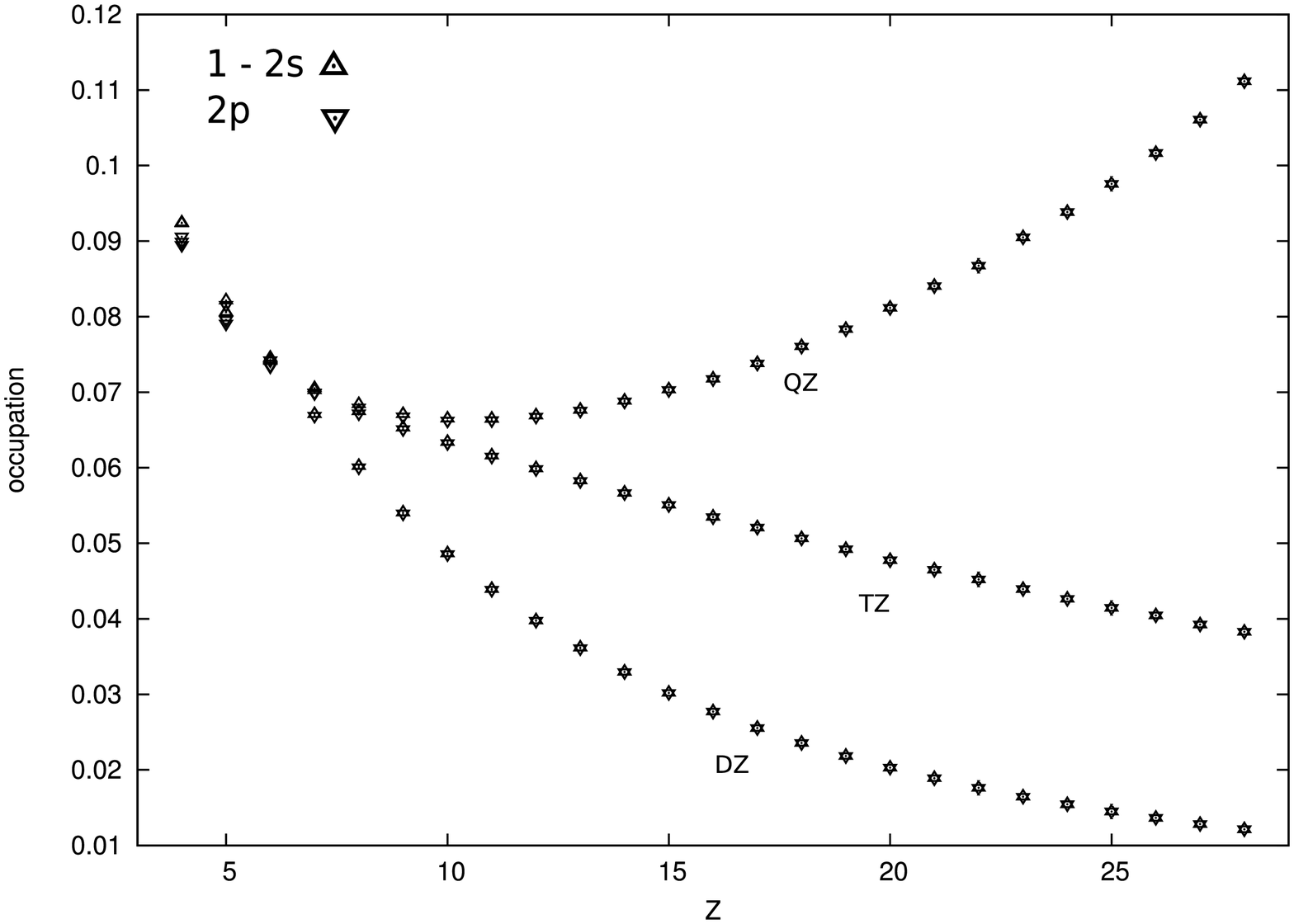}
\caption{\label{nocc_2s2p} The natural occupation of the $2p$-orbital and one minus the occupation of $2s$-orbital for the Be series, in all three basis sets.}
\end{figure}
The eigenvalues of the sp density matrix (i.e.\ the natural orbital occupation numbers) provide insight into the extent of correlation.  
The occupation numbers from SDP are always very close to those from full-CI, differing by at most 0.005. 
Of particular interest are the occupations of the quasidegenerate $2s$ and $2p$ orbitals in Be. These are shown in Fig.~\ref{nocc_2s2p}. 
The sum of the $2s$ and $2p$ occupations is nearly 1 and increasingly so for large $Z$. This implies that only the $2s$ and $2p$ are 
partially occupied in the large-$Z$ limit. The shapes of the curves reflect the aforementioned imperfections in the basis sets, 
with the $2s$ below the $2p$ for cc-pVDZ and cc-pVTZ, and above the $2p$ for cc-pVQZ.
\section{Summary}
%abstract
Variational methods based on the second-order density matrix seem to hold great promise as an ab initio many-body technique, but there is room for improvement , especially as regards computational efficiency (improved algorithms) and accuracy (better characterization of the $N$-representable set). We investigated the isoelectronic series of Be, Ne and Si using the P, Q and G  $N$-representability conditions. A significant speedup is obtained when spin and rotational symmetry is taken into account. This allowed us to investigate the properties of the SDP method with increasing basis set size (cc-pV(D,T,Q)Z). The energies so obtained are reasonably accurate, but the accuracy seems to diminish with increasing basis set size. The SDP method is capable of describing the strong static electron correlations appearing in the beryllium and silicon series due to the incipient degeneracy in the hydrogenic spectrum for increasing central charge. The ionization energies, constructed using the extended Koopmans' theorem, are surprisingly good. Also the natural occupations are reproduced very well when compared to full-CI results in the same basis sets. Hence, the physical content of the variationally determined second-order density matrix seems to be reliable. 

Apart from a study of the potential energy surface for some diatomic 14-electron molecules \cite{helen_2}, we intend to investigate fermionic and bosonic Hubbard models on one and two - dimensional lattices. Further work is also needed to ameliorate the computational cost of the method, and to increase the accuracy without introducing three-index conditions.
\appendix*
\section{Constraint matrices}
The linear constraints for imposing the spin singlet condition are given by:
\begin{equation}
\forall ~~ k \leq l \qquad:\qquad\mathrm{Tr}~\Gamma~ ^{[kl]}K= 0~, 
\end{equation}
where the constraint matrices $ ^{[kl]}K $ have the following form
\begin{eqnarray}
\nonumber^{[kl]}K^{S}_{ab;cd} &=& ^{[kl]}f^{S}_{ab;cd} + (-1)^S {}^{[kl]}f^{S}_{ba;cd} + (-1)^S {}^{[kl]}f^{S}_{ab;dc}+ {}^{[kl]}f^{S}_{ba;dc} \\
&&+ {}^{[kl]}f^{S}_{cd;ab} + (-1)^S {}^{[kl]}f^{S}_{dc;ab}  + (-1)^S {}^{[kl]}f^{S}_{cd;ba} + {}^{[kl]}f^{S}_{dc;ba}~,
\end{eqnarray}
with
\begin{equation}
^{[kl]}f^{S}_{ab;cd} = \left[\frac{1}{2}\frac{1}{N-1} - (-1)^S\left\{\begin{matrix}\frac{1}{2} & \frac{1}{2} & 1 \\ \frac{1}{2} & \frac{1}{2} & S\end{matrix}\right\}\right]\delta_{ak}\delta_{cl}\delta_{bd}~.
\end{equation}
The constraints for spin and angular momentum singlet projection are:
\begin{equation}
\forall ~~ k \leq l \qquad:\qquad\mathrm{Tr}~\Gamma~ ^{[kl]}_{\mathcal{J}}K= 0~,
\end{equation}
where the constraint matrices $ ^{[kl]}K $ have the following form
\begin{eqnarray}
\nonumber^{[kl]}_{~~\mathcal{J}}K^{(L^\pi S)}_{ab;cd} &=& {}^{[kl]}_{~~\mathcal{J}}f^{(L^\pi S)}_{ab;cd} + (-1)^{L+S+l_a+l_b} {}^{[kl]}_{~~\mathcal{J}}f^{L^\pi S}_{ba;cd} + {}^{[kl]}_{~~\mathcal{J}}f^{(L^\pi S)}_{ba;dc} + (-1)^{S + L + l_a + l_b} {}^{[kl]}_{~~\mathcal{J}}f^{(L^\pi S)}_{ab;dc}\\
&& + {}^{[kl]}_{~~\mathcal{J}}f^{L^\pi S}_{cd;ab} + (-1)^{S+L+l_a+l_b} {}^{[kl]}_{~~\mathcal{J}}f^{L^\pi S}_{dc;ab} + {}^{[kl]}_{~~\mathcal{J}}f^{L^\pi S}_{dc;ba} + (-1)^{L+S+l_a+l_b} {}^{[kl]}_{~~\mathcal{J}}f^{L^\pi S}_{cd;ba}~,
\end{eqnarray}
where $\mathcal{J}$ can mean either $\Sigma$ or $\Lambda$ and
\begin{eqnarray}
\nonumber ^{[kl]}_{~~\Sigma}f^{(L^\pi S)}_{ab;cd} &=& \delta_{l_kl_l}\left[\frac{1}{2}\frac{1}{N-1} - \frac{(-1)^S}{[l_k]}\left\{\begin{matrix}\frac{1}{2} & \frac{1}{2} & 1 \\ \frac{1}{2} & \frac{1}{2} & S\end{matrix}\right\}\right]\\
&& \times\delta_{ak}\delta_{cl}\delta_{bd}~,\\
\nonumber ^{[kl]}_{~~\Lambda}f^{(L^\pi S)}_{ab;cd} &=& \delta_{l_kl_l}\left[\frac{\hat{l}^2_a}{N-1} - \frac{1}{2}\left(\hat{l}_a^2 + \hat{l}_b^2 - \hat{L}^2\right)\right]\\
&& \times\delta_{ak}\delta_{cl}\delta_{bd}~.
\end{eqnarray}
\begin{acknowledgments}
We would like to thank ir. Matthias Degroote for the CCSD and full-CI results. We gratefully acknowledge financial support from FWO-Flanders and the research council of Ghent University. PWA acknowledges support from NSERC and Sharcnet. B.V., H.V.A., P.B. and D.V.N. are Members of the QCMM alliance Ghent-Brussels.
\end{acknowledgments}
\bibliography{atomic.bib}
\end{document}